# Green, red and IR frequency comb line generation from single IR pump in AlN microring resonator

Hojoong Jung,[1] Rebecca Stoll,[2] Xiang Guo,[1] Debra Fischer[2] and Hong X. Tang[1,*]

[1]*Department of Electrical Engineering, Yale University, New Haven, Connecticut 06511, USA*
[2]*Department of Astronomy, Yale University, New Haven, Connecticut 06511, USA*
*Corresponding author: hong.tang@yale.edu*

On-chip frequency comb generations enable compact broadband sources for spectroscopic sensing and precision spectroscopy. Recent microcomb studies focus on infrared spectral regime and have difficulty in accessing visible regime. Here, we demonstrate comb-like visible frequency line generation through second, third harmonic, and sum frequency conversion of a Kerr comb within a high Q aluminum nitride microring resonator pumped by a single telecom laser. The strong power enhancement, in conjunction with the unique combination of Pockels ($\chi^2$) and Kerr ($\chi^3$) optical nonlinearity of aluminum nitride, leads to cascaded frequency conversions in the visible spectrum. High-resolution spectroscopic study of the visible frequency lines indicates matched free spectrum range over all the bands. This frequency doubling and tripling effect in a single microcomb structure offers great potential for comb spectroscopy and self-referencing comb.

Optical frequency combs consist of equally spaced frequency components and can be generated by cascaded four-wave mixing (FWM) in Kerr nonlinearity materials with dense optical power [1, 2]. High quality (Q) factor microresonators have been used to enhance the optical power and generate the frequency comb in various Kerr materials, such as silica [3, 4], high-index silica-glass [5], crystalline $CaF_2$ [6, 7], $MgF_2$ [8, 9], silicon nitride (SiN) [10-12], and aluminum nitride (AlN) [13]. Their bandwidth is essentially controlled by phase matching condition of FWM in a microring. On the other hands, frequency combs from femtosecond lasers have been used to produce frequency components in shorter wavelength region by wave mixings in noble gas (Xe) [14] and $\chi^{(2)}$ crystals (such as PPKTP) [15]. However, femtosecond laser based combs require sophisticated setup that involves separate cavity and nonlinear crystals. Here, we leverage resonant enhanced wave mixing in microcavities and report the generation of frequency lines in IR, red and green regions using an AlN microring resonator.

Second harmonic generation (SHG) in whispering gallery mode resonators has been reported in various $\chi^{(2)}$ materials and structures [16-20]. Third harmonic generation (THG) also has been demonstrated in silica micro-toroid [21] and SiN microring [22] with their strong $\chi^{(3)}$ nonlinearity. Generally speaking, centrosymmetric materials show the third-order nonlinear effects, but little $\chi^{(2)}$ effects. In certain cases, the material centrosymmetry can be broken for instance at the waveguide interface of SiN and as a result, SHG is observed by the effective $\chi^{(2)}$ nonlinearity in addition to the THG [22]. Multiple frequency lines in both the infrared (IR) and visible range using a single IR CW pump is reported by further investigation of SHG and FWM in SiN microring resonator [23]. However, to attain stronger $\chi^{(2)}$ nonlinearity, materials having strong intrinsic Pockels effect are more desired as the strength of surface Pockels effect can be limited. AlN, on the other hand, has intrinsic Pockels effect in addition to Kerr nonlinearity [24, 25].

Here we demonstrate the same spacing frequency lines from green to IR regions via simultaneous $\chi^{(2)}$ and $\chi^{(3)}$ effects in AlN microring. Green frequency lines near 517 nm and red lines near 776 nm are measured using CHIRON high resolution spectroscopy [26, 27]. By analyzing the spectrum, we found that within our resolution limit all visible and IR combs show the identical frequency gap which is the free spectral range (FSR) of microring in IR region. The doubled and tripled frequency lines in the ring can be used either on-chip $f$-2$f$ (or $2f$-$3f$) self-referencing, once octave (or 2/3) octave IR comb is achieved. This can dramatically simplify the comb generating system by eliminating additional nonlinear materials and free-space optical components for comb self-referencing and stabilization.

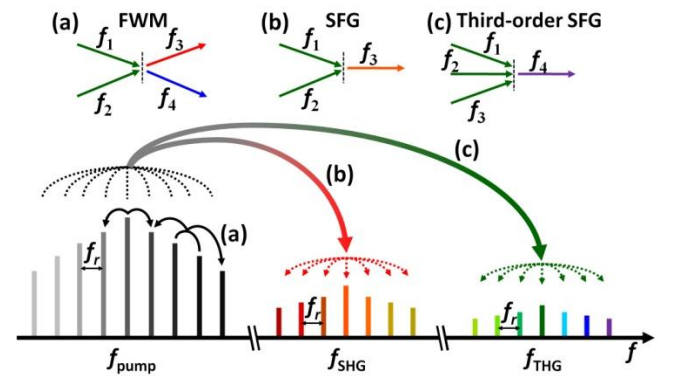

Fig. 1 The principle of multiple frequency conversions through FWM, SFG and third-order SFG processes. (a) Cascaded FWM generates the fundamental comb near the IR pump. (b) Through SHG and SFG, IR comb lines can combine and generate other frequency lines at the doubled frequency of the pump with the same reputation rate ($f_r$) as the IR comb. (c) Three IR comb lines can combine and generate frequency lines at the triple of IR pump frequency through third-order SFG.

Figure 1 illustrates the principle of multi-frequency lines generation through three-wave mixing (TWM) and FWM processes. In a high $Q$ microring resonator, we can observe a Kerr frequency comb near the pump frequency when the circulating power in the microring is beyond a threshold (Fig. 1 (a)) that can be expressed by $f_{IR} = f_o + nf_r$, where $f_o$ is the offset frequency, $f_r$ is the comb repetition which is determined by the FSR of the microring, and $n$ is an integer. Frequency mixing among the comb lines (SHG and SFG) through Pockels effect (Fig. 1 (b)) creates frequency lines with the same $f_r$ as the IR comb at twice of the IR pump frequency. This frequency lines in red region can be expressed by,

$$f_{red} = f_{IR}^1 + f_{IR}^2 = 2f_o + (n_1 + n_2)f_r = 2f_o + mf_r \quad (1)$$

Because $m$ is an integer, the IR comb can be duplicated in red region, albeit at twice offset frequency. Similar to the SHG and second-order SFG case, frequency lines near the triple of IR pump frequency can be generated by THG and third-order SFG (Fig. 1(c)),

$$f_{green} = f_{IR}^1 + f_{IR}^2 + f_{IR}^3 = 3f_o + lf_r \quad (2)$$

By this combination of second and third order optical nonlinear effects, we can generate IR comb by FWM and upconvert it to red and green region by second and third-order SFG. In addition, there could be also other processes taking effect, such as FWM between IR pump and SHG lines can generate IR and red frequency lines, or SFG between IR and red lines to generate green lines. However, the direct conversion from IR comb to red and green lines are dominant processes due to the high circulating power of IR comb and non-resonant SHG peak.

Phase matching condition should be satisfied for efficient wavelength conversion. For SFG, energy and momentum conservations are $f_1 + f_2 = f_3$ and $n_1 f_1 + n_2 f_2 = n_3 f_3$, respectively. To satisfy these equations, refractive index $n_3$ should be between $n_1$ and $n_2$. Third-order SFG case is the same with one more term. Figure 2 shows the simulated effective indices for a waveguide of 650 nm by 3.5 μm, which represents the cross section of our microring resonator. The black solid line is the index of fundamental TE-like mode near 1550 nm vacuum wavelength region. The red and green lines are the indices near 775 and 520 nm wavelength regions, respectively. Two modes in red region and 11 modes in green region cross the black solid line. Some modes are indicated as examples. If their mode overlap (3) is non-zero (in here 0.047 and 0.007 for red region, and from 0.0035 to 0.44 for green region), phase matching

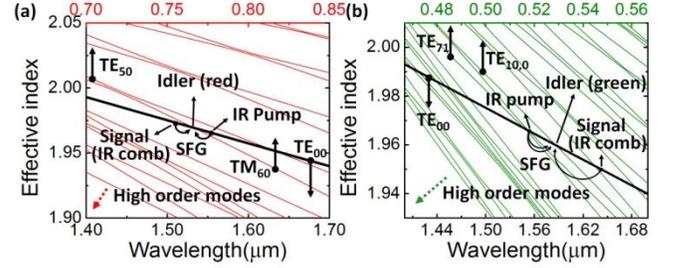

Fig. 2. The simulated effective indices near 1550 nm (black lines in (a) and (b), TE$_{00}$), 775 nm (red lines in (a)) and 517 nm wavelength (green lines in (b)) of AlN microring. Phase matching windows are near each cross points of black and red lines (or green lines for third-order SFG).

conditions can be satisfied near there. The mode overlap is defined as:

$$\xi = \frac{2\pi R}{\left[(\frac{\hbar\omega_a}{\varepsilon_0 \varepsilon_a V})^n \frac{\hbar\omega_b}{\varepsilon_0 \varepsilon_b V}\right]^{1/2}} \int_{Waveguide} dxdy (u_{a,x}^*)^n u_{b,x} \quad (3)$$

where $R$ is the microring radius, $V$ is the mode volume, $\omega_a$, $\varepsilon_a$, $\omega_b$ and $\varepsilon_b$ are the fundamental frequency, its permittivity, SHG (THG) frequency and its permittivity, respectively. $n$ is the number of pump photons, for example, $n = 2$ for SHG and $n = 3$ for THG. $u_{a,x}$ and $u_{b,x}$ are the electric field along $x$ direction (parallel to the plane) for fundamental and SHG (THG) wavelength. The mode overlaps in here are all non-zero due to the waveguide bending of microring.

In ring resonators, the IR and visible resonant modes are discrete, and their center resonances rarely commensurate with each other in as-fabricated devices. They can be brought to match through thermal tuning of resonant wavelengths as optical power in the device is increased. Since the quality factor of the visible modes are often much lower than that of the IR mode, there is a reasonable margin to achieve wavelength matching. In the case of SFG, the best phase matched combination of signal and pump can be chosen among the IR comb peaks, so that their SFG power can be comparable to THG power. Although the mode coupling between fundamental mode to higher order modes are not strong, the high circulating intensity in the ring made the conversion still quite observable. For practical purpose, the fundamental to fundamental mode conversion is more desired to attain maximum modal overlap. This can be achieved by periodic poling [16], birefringent materials index tuning [17], or using photonic crystal structure [28].

The fabrication starts with the sputtering of the highly c-axis oriented, 650 nm AlN on a 2 μm-thick thermal SiO$_2$ wafer using an S-gun magnetron sputtering system. Microrings of 3.5 μm in width and 60 μm in radius are then patterned to the AlN film with coupling bus waveguides using E-beam lithography. The chip is then covered with 3.5 μm-thick plasma-enhanced chemical vapor deposition (PECVD) $SiO_2$. Following thermal annealing at 1000 $^0$C for 30 hours in an $O_2$ atmosphere, the device is diced out and polished for fiber butt coupling. The left inset of Fig. 3 (a) shows the top view of the microring, and right inset is a cross-sectional SEM image of the AlN waveguide. For efficient and mechanically stable fiber to chip coupling, 4 μm mode field diameter (MFD) single mode fibers are touched to the chip at both input and output sides. 40 % coupling efficiency is achieved per facet, and there is no observable mechanical drift during the measurement time.

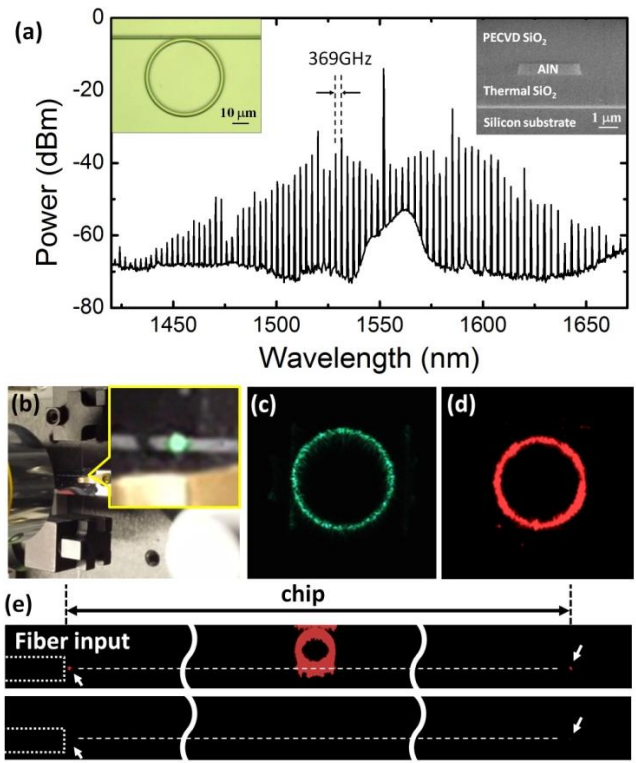

Fig. 3. (a) IR frequency comb generated in 60 μm radius microring resonator (left inset) with 650 nm height and 3.5 μm width structure (right inset). Estimated power in the bus waveguide is 600 mW and 0.5 % of gathered output power is used for spectrum measurement. (b) The greenly glowing microring under daylight. (c) The microscope image without a filter and (d) with a 750 nm long pass filter. (e) The microscope image with the 750 nm long pass filter when on-resonance (top) and off-resonance (bottom). The white dashed lines represent the bus waveguides and the white dotted boxes on the left represent the input fibers. The white arrows indicate the ends of the bus waveguides of the chip.

To generate IR Kerr comb from this device, a tunable continuous wave (CW) diode laser and erbium-doped fiber amplifier (EDFA) are used to create 3W CW pump. Then we scan the pump wavelength from shorter to longer wavelengths and achieve "thermal lock" on a high $Q$ optical resonance [29]. Due to the enhanced circulating power in the microring, frequency comb can be generated via FWM and is observed using the optical spectrum analyzer (OSA) through the output fiber as shown in Fig. 3 (a). The threshold power in the bus waveguide is estimated about 300 mW and the frequency repetition rate is 369 GHz which corresponds to the fundamental mode FSR of microring at 1550 nm wavelength. The comb span bandwidth is around 250 nm which is limited by dispersion [2]. We note that the dispersion can be further compensated by using thicker AlN films or by engineering the sidewall profile of etched waveguide. The loaded $Q$ factor of the microring for this comb generation is about 500,000. We expect that the optical $Q$ can be further improved by optimizing fabrication process such as resist reflowing to reduce roughness of etched surface and also by using a thicker film to better confine the light and reduce the surface absorption. Once the IR comb is generated in the microring, visible frequency lines are also generated through TWM and FWM. First, we confirm the strong wavelength conversion from IR to visible wavelength in the microring by observation of the glowing ring. Obviously, the green light can be observed with bare eyes even with daylights as shown in the inset of Fig. 3 (b). Figure 3 (c) shows the greenly shining ring resonator taken with a visible CCD camera. Due to the high sensitivity of green light to human eyes and relative insensitivity of our imaging setup in visible range of red frequency lines (center wavelength of 776 nm), only green color is observed in Fig 3 (b) and (c). To confirm the IR to red wavelength conversion in the microring, green light is blocked by a 750 nm long pass filter. Then the microscope image of a red ring resonator can be obtained (Fig. 3 (d)).

Before we analyze the output light with the CHIRON high-resolution spectrometer, we further check that the upconverted light is indeed from the ring resonator, not from bus waveguide. Figure 3 (e) shows microscope images after the 750 nm long pass filter when on-resonance (top) and off-resonance (bottom) with 20s exposure time. The white dashed lines are guide lines representing bus waveguides. The IR pump laser is injected from the left input fiber expressed by white dotted box. On resonance, strong red light is generated in the ring resonator and some of the light couples out to the bus waveguide. At the ends of the bus waveguide, bright spots can be observed at input and output facets due to the clock wise (CW) and counter clock wise (CCW) modes in the ring resonator [30]. In the case of off-resonance, however, there is a very weak spot at the output facet only, indicative of SHG from propagating IR pump light in the bus waveguide. This can be ignored compared to the strong circulating lights in the ring resonator when it is tuned on-resonance.

Figure 4 (a) is the simplified schematics of our CHIRON setup for visible spectrum measurement. IR, red and green lights are generated in the microring as explained above, and are coupled into an output single mode fiber. 0.5 % of collected power is sent to a power meter for fiber to chip feedback alignment, and 0.5 % is sent to the OSA for IR comb monitoring. The remaining 99 % of output power is sent to the spectrometer for spectral characterization in visible range. In the spectrometer, an Echelle grating is used as the dispersive element by dividing the incoming light into discrete orders vertically depends on the frequency. A following prism provides the cross-dispersion by spreading the spectrum horizontally. Finally, a two dimensional spectrum is obtained at the CCD camera with high resolution. Note that each row covers 4 nm (at 410 nm) to 13 nm (at 880 nm) wavelength span and a total of 73 rows span from 410 nm to 880 nm wavelength.

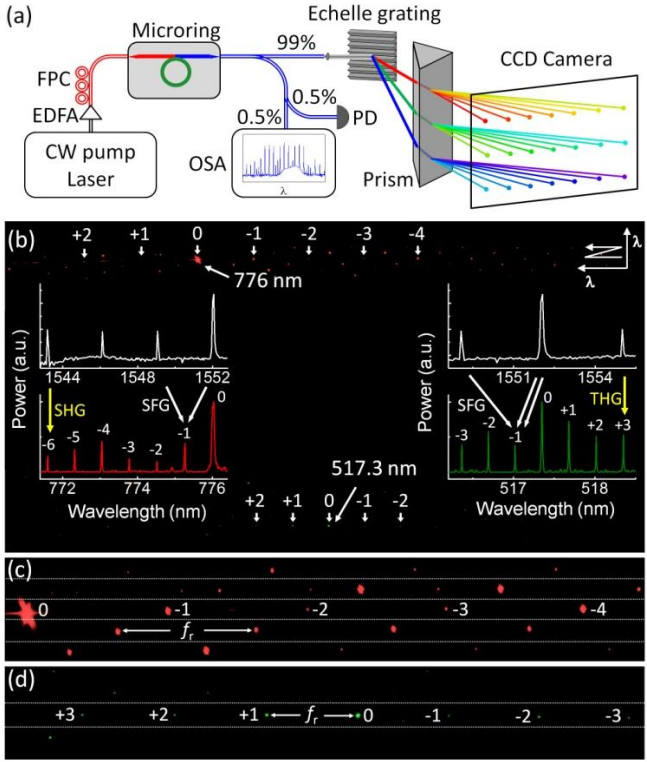

Fig. 4. (a) The simplified schematics of experimental set up for comb generation and spectrum measurement. CW: continuous wave, EDFA: erbium-doped fiber amplifier, FPC: fiber polarization controller, PD: photo detector, OSA: optical spectrum analyzer. (b) The portion of detected image from the CCD camera and the extracted spectra from it (insets). White and yellow arrows in inset spectra show some examples of SFG, SHG and THG. (c) A zoom-in image of 776 nm and (d) 517 nm spectral region. The white dotted lines represent the boundary between different grating orders.

Figure 4 (b) is a portion of CCD image obtained with 20 min exposure time when the pump wavelength is tuned to 1552 nm resonance. The SHG peak at 776 nm wavelength has 1.2 nW estimated power while SFG peaks between 740 nm and 800 nm have powers varying from 6 pW to 75 pW. The power of THG peak at 517.3 nm wavelength is ~ 51 pW and third-order SFG peaks between 510 nm and 540 nm are from 1 pW to 17 pW. All these powers refer to the power in the output fiber. The net efficiency detected from the output waveguide is about $10^{-8}$ for SHG and $10^{-9}$ for THG in visible region. The actual conversion efficiency in the ring should be much higher due to the low coupling from microring to bus waveguide which can be improved by adding a separate drop port waveguide that is specifically designed for target extraction wavelength in the visible regime [19]. Total 84 and 43 measurable peaks are detected near 776 nm and 517.3 nm, respectively, although they are not all shown in the Fig. 4 (b) due to the limited figure resolution. Some peaks near the 776 nm and 517 nm are numbered to track them in spectra and zoom-in images. The top insets show the spectra measured by the OSA, and the bottom insets are extracted spectra from Fig. 4 (b) with normalized power. In the left inset, seven red frequency lines generated by SFG and SHG are labeled. The green frequency lines arising from THG and third-order SFG are shown in the right inset with corresponding identification labels.

A zoom-in image of 776 nm and 517 nm spectral regions are shown in Fig. 4 (c) and (d), respectively. The white dotted lines represent the boundary between different grating orders. Within each grating orders, spots are identified with a spacing of 369.2 ± 0.2 GHz, which matches mode spacing ($f_r$) in IR comb. SHG and THG from IR comb produce only frequency lines with $2f_r$ and $3f_r$ spacing, respectively. Therefore, the remaining lines are from the SFG of IR frequencies (Fig. 1(b), (c)). By this combination of different nonlinear effects, the same $f_r$ is maintained among all the spectral regimes. The inherent dispersion in CHIRON spectrometer makes the physical spacing appear different between Fig. 4 (c) and Fig. 4 (d), but they are all the same in frequency domain.

In conclusion, three frequency comb-like sets from IR to red to green are generated in AlN microring resonator from a combination of TWM and FWM pumped by single telecom IR pump laser. The spectrum of visible portion is measured using CHIRON high-resolution spectrometer and confirmed to have the same frequency gap. Our microcomb benefits from the co-existing $\chi^{(2)}$ and $\chi^{(3)}$ nonlinearity of AlN, the high finesse as well as the high modal confinement in the micro-ring resonator. This unique property of AlN microring has great potentials in wideband comb generation from visible to IR. Once this is combined with octave or 2/3 octave IR comb, on-chip nonlinear interferometry self-referencing for comb stabilization is possible.

This work was supported by a Defense Advanced Research Projects Agency (DARPA) grant under its PULSE program. H.X.T. acknowledges support from a Packard Fellowship in Science and Engineering and a National Science Foundation CAREER award. Facilities used were supported by Yale Institute for Nanoscience and Quantum Engineering and NSF MRSEC DMR 1119826. The authors thank Michael Power and Dr. Michael Rooks for assistance in device fabrication.